\documentclass[apj]{emulateapj}

\newcommand\markcite[1]{}

\gdef\h50min{$h_{50}^{-1}$}

\gdef\3727{[O\,{\sc ii}]\,3727\,\AA}
\gdef\o4959{[O\,{\sc iii}]\,4959\,\AA}
\gdef\5007{$\lambda \lambda 5007$\,[O\,{\sc iii}]}

\gdef\4ang{4000\,\AA}

\gdef\pegase{P\'{E}GASE}
\gdef\eazy{\textsc{EAZY}}
\gdef\zspec{z_\mathrm{spec}}
\gdef\zphot{z_\mathrm{phot}}
\gdef\zml{z_\mathrm{ML}}
\gdef\zprior{z_\mathrm{prior}}
\lefthead{Brammer, van Dokkum \& Coppi}
\righthead{EAZY}
\slugcomment{Draft; version \today}
\begin{document}

\title{EAZY: A Fast, Public Photometric Redshift Code}

\slugcomment{Accepted to The Astrophysical Journal}

\author{Gabriel~B.~Brammer\altaffilmark{1},
Pieter~G.~van Dokkum\altaffilmark{1}, and Paolo Coppi\altaffilmark{1}
}

\altaffiltext{1}{Department of Astronomy, Yale University,
New Haven, CT 06520-8101}

\begin{abstract}

We describe a new program for determining photometric redshifts,
dubbed EAZY. The program is
optimized for cases where spectroscopic redshifts are not
available, or only available for a biased subset of the galaxies.
The code combines features from various existing codes: it can
fit linear combinations of templates, it includes optional
flux- and redshift-based priors, and its user interface is
modeled on the popular HYPERZ code.
A novel feature is that the default template set, as well as
the default functional forms of the priors, are not based on
(usually highly biased) spectroscopic samples, but on
semi-analytical models.
Furthermore, template mismatch is
addressed by a novel rest-frame template error function. This
function gives different wavelength regions different weights,
and ensures that the formal redshift uncertainties are realistic.
We introduce a redshift quality parameter, $Q_z$,
that provides a robust estimate of the reliability of
the photometric redshift estimate.
Despite the fact that EAZY is not "trained" on
spectroscopic samples, the code (with default parameters) performs
very well on existing public datasets. For $K$-selected
samples in CDF-South and other deep fields
we find a $1\sigma$ scatter in $\Delta z/(1+z)$ of 0.034,
and we provide updated photometric
redshift catalogs for the FIRES, MUSYC, and FIREWORKS surveys.

\end{abstract}

\keywords{cosmology: observations ---
galaxies: evolution --- galaxies:
formation
}

\section{Introduction}

Accurate redshifts of distant galaxies are crucial for nearly all of
observational cosmology. Whereas extensive spectroscopy with multi-object
spectrographs on 8-10m class telescopes has yielded redshifts for thousands,
and in some cases tens of thousands, of galaxies
\citep[e.g.][]{steidel:03,deep2,vvds}, these galaxies tend to be relatively bright at
optical wavelengths. For galaxies fainter than $R\sim 25$ we rely almost
exclusively on photometric redshifts, derived from fitting template spectra to
broad- or medium-band photometry
\citep[e.g.][]{lanzetta:96,wolf:03,franx:03,mobasher:04,drory:05}. This
situation is not likely to change, even with the advent of efficient
spectrographs with very wide fields \citep[such as WFMOS;][]{wfmos},
multi-object capabilities in the near-infrared \citep[e.g. MOIRCS;][]{moircs},
or larger telescopes. The signal-to-noise ratio (S/N) per resolution element in
the continuum decreases with spectral resolution as ${\rm S}/{\rm N} \propto
R^{-0.5}$ for a given exposure time. Therefore, the required integration time
to maintain a given S/N per resolution element increases linearly with the
spectral resolution, quite independent of the details of the telescope and
instruments. As a typical set of broad band filters corresponds to $R\sim 5$
and typical faint object spectrographs have $R\sim 1000$, spectroscopy is about
two orders of magnitude more time consuming than photometry for a given
telescope size. A notable exception is spectroscopy of emission line objects,
which can be extremely efficient at faint magnitudes. 

The methodology for determining photometric redshifts using the template-fitting
approach is essentially straightforward: the photometric data are compared to synthetic
photometry for a large range of template spectra and redshifts, and the most likely
redshift follows from a statistical analysis of the differences between observed and
synthetic data. Several codes exist that perform this task, each employing its own
techniques for creating the synthetic photometry and interpreting the residuals in the
redshift -- template plane. Popular examples include HYPERZ
\citep{hyperz}, \textsc{ImpZ}
\citep{impz}, and Le PHARE\footnote{\texttt{http://www.oamp.fr/people/arnouts/LE\_PHARE.html}} (Arnouts \&
Ilbert), which do a straightforward $\chi^2$ minimization; GREGZ\footnote{Greg Rudnick
did not name his code; the name GREGZ is used for convenience in the present paper.}
\citep{rudnick:01,rudnick:03}, which allows linear
combinations of templates and uses Monte Carlo methods to determine the redshift
uncertainties; BPZ \citep{bpz}, which uses Bayesian
statistics allowing the use of priors; and ZEBRA \citep{zebra} and \texttt{kcorrect} \citep[][hereafter BR07]{blanton:07}, which include (distinct) iterative template-optimization routines that make use of the extensive spectroscopic databases of the zCOSMOS \citep{lilly:07} and Sloane Digital Sky Survey \citep[SDSS; ][]{york:00} projects, respectively.

For obvious reasons photometric redshifts benefit from having high quality
photometry in many bandpasses and from sampling strong continuum features in
the observed wavelength region (such as a Lyman or Balmer break), irrespective
of the methodology. However, given a set of objects with good quality
photometry, the aspect that is of paramount importance for obtaining reliable
photometric redshifts is the selection of the template set (see Feldmann et
al.\ 2006, \S\ref{s:template_set}). \cite{zebra} obtain very good results by
iteratively adapting the templates, minimizing the systematic differences
between the best fitting templates and the actual galaxy photometry. This
approach not only reduces the random uncertainty in the photometric redshifts
but can also eliminate systematic effects in certain redshift ranges
\citep[see][]{zebra}. The disadvantage of this optimization is that its effects
can only be assessed when a large sample of galaxies with spectroscopic
redshifts is available, and when this sample is a random subset of the entire
photometric sample. This assumption may be valid in the case of zCOSMOS, but
this is generally not the case in studies of galaxy samples which are
significantly fainter than the spectroscopic limit.

In this paper we describe a new photometric redshift code which was written
specifically for samples with incomplete and/or biased spectroscopic
information (such as, for example, faint $K$-selected samples).
Rather than minimizing the scatter in the relation between
photometric and spectroscopic redshift using the spectroscopic sample as a
training set, a user-defined template error function is introduced to account
for wavelength-dependent template mismatch. The code combines features from
various existing programs: the possibility of fitting linear combinations of
templates (as done in GREGZ), the use of priors (as first done in BPZ), and a
user-friendly interface based on HYPERZ. The default template
set and the redshift-magnitude priors are derived from semi-analytical models. These
models are, of course, only an approximation of reality, but their ``perfect''
completeness down to very faint magnitudes outweighs their imperfect
representation of the real Universe. 

The outline of this paper is as follows.  In \S\ref{s:implementation}
we describe the implementation of the code, including the optimized
template set and redshift priors derived from semi-analytical models
and the template error function derived from the GOODS-CDFS
photometric catalog.  In \S\ref{s:application} we test the code on a
combined photometric catalog from a variety of deep multi-wavelength
surveys and compare the photometric redshifts to spectroscopic
redshifts of nearly 2000 galaxies at $0 < z < 4$.  In
\S\ref{s:reliability} we discuss the reliability of the photometric
redshift estimates and provide cautionary examples for relying solely
on spectroscopic samples to estimate the photometric redshift quality.
Finally, in \S\ref{s:summary} we summarize the features and
performance of the photometric redshift code and discuss future avenues
for improvement.

\section{Implementation}
\label{s:implementation}

\subsection{Basic Algorithm}
\label{s:basic}

The basic algorithm is similar to many existing photometric redshift codes. The algorithm steps through a user-defined grid of redshifts, and at each redshift finds the best fitting synthetic template spectrum by minimizing

\begin{equation}
\chi^2_{z,i} = \displaystyle\sum_{j=1}^{N_{\rm filt}} \frac{\left( T_{z,i,j}
- F_{j}\right)^2}{\left( \delta F_{j} \right)^2},
\label{eq:chi2}
\end{equation}

\noindent with $N_{\rm filt}$ the number of filters, $T_{z,i,j}$ the synthetic
flux of template $i$ in filter $j$ for redshift $z$, $F_j$ the observed flux in
filter $j$, and $\delta F_{j}$ the uncertainty in $F_j$. Templates are
corrected for absorption by intervening \ion{H}{1} clouds following the
\cite{madau:95} prescription. The template fit is done in linear space, as
this allows a proper treatment of flux errors and of negative
flux measurements.

In most photometric redshift codes (e.g. HYPERZ, BPZ, ZEBRA) each template
$T_i$ is a single-component empirical or synthetic spectral energy distribution
(SED). However, in practice many galaxies are not well represented by any
individual template from the user-supplied library, and as a result template
mismatch is the primary source of error in photometric redshift estimates. In
ZEBRA, the detailed form of each template is adapted iteratively based on
residuals from fits to galaxies with spectroscopic redshifts. Instead, we
follow GREGZ and allow linear combinations of templates. Rather than finding
the best-fitting template $T_i$ the code finds the best-fitting coefficients,
$\alpha_i$, in

\begin{equation}
T_{z} = \displaystyle\sum_{i=1}^{N_{\rm temp}} \alpha_i T_{z,i},
\label{eq:t_i}
\end{equation}

\noindent with all $\alpha_i\geq 0$. The number of
template components fit simultaneously is one, two, or all
of the templates in a user-defined list. 
For the one- and two-template fits the coefficients, $\alpha_i$, are determined
using analytic least-squares fits, while for the latter option the coefficients
for every template in the library are determined iteratively following the
algorithm of \cite{sha_nmf:07}. In practice, the choice between a 2-template
fit and an $N$-template fit is a trade-off between accuracy and computation
time. The improvement going from two templates to $N$ templates can be
significant if the number of templates in the library is small ($\lesssim 5$),
but is usually negligible when the number of templates is large ($\gtrsim 10$).

\begin{figure}
\plotone{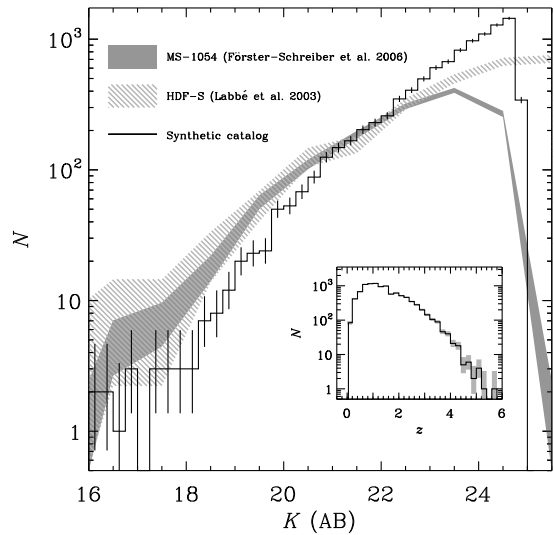}
\figcaption{Distribution of $K$ magnitudes of $10^4$ galaxies from our
simulated lightcone catalog.  The $K$-band number counts of the two deep fields
from the FIRES survey \citep{labbe:03,forster:06} are shown for comparison,
scaled by the ratio of the number of galaxies with $K_s<23.5$ in these fields
compared to the lightcone catalog.\textit{inset:} Redshift distribution of the
galaxies in the lightcone catalog.  The widths of the shaded FIRES
distributions and the error bars on the histograms are the poisson-like errors
for small numbers calculated following \cite{gehrels:86}.  The simulated number
counts are generally consistent with the observed values to within a factor of
$\sim$2.  The simulated catalog provides a unique template calibration set that
is complete at the limiting magnitudes characteristic of deep imaging surveys,
and that includes thousands of galaxies at $z>2$ where observed spectroscopic
samples are sparse. \label{fig:lc_dist}}
\end{figure}

\subsection{Optimized template set}
\label{s:template_set}

\begin{figure*}
\plotone{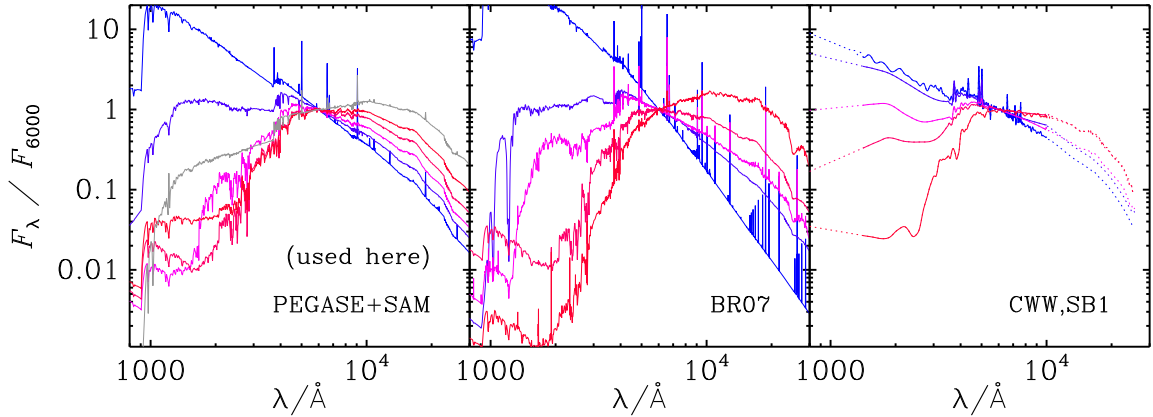}
\figcaption{\textit{left:} Five templates generated following the \cite{blanton:07} algorithm with \pegase\ models and a calibration set of synthetic photometry derived from semi-analytic models.  Shown in grey is the additional young and dusty template added to compensate for the lack of extremely dusty galaxies in the SAMs.   All of the templates are shown normalized at 6000\AA.  \textit{middle:} ``Default'' template set of \cite{blanton:07}.  \textit{right:} Empirical templates from \cite{cww} plus the ``SB1'' starburst spectrum from \cite{kinney:96} that are frequently used for photometric redshifts.  The templates shown are extended into the NUV and NIR with \cite{bc:03} models \citep[dashed regions; see, e.g.][]{rudnick:01}.  \label{fig:show_templates}}
\end{figure*}

The recent significant improvement in the quality of maximum-likelihood
photometric redshift estimates has come largely from optimizing the template
set that is fit to the broad-band photometry.  In practice, the template set
needs to be large enough that it spans the broad range of multi-band galaxy
colors and small enough that the color and redshift degeneracies are kept to a
minimum \citep[e.g.][ ]{bpz}.  The combined set of empirical galaxy templates from \cite{cww} and 
\cite{kinney:96} (hereafter, CK) is the
most frequently used set for photometric redshift measurements \citep[BPZ; ZEBRA; ][]{ilbert:06,mobasher:07}, and it provides the additional benefit
of allowing a rough estimate of an individual galaxy's spectral type as well as
its redshift.  The CK set suffers from a number of disadvantages,
however: it is determined from local galaxies and is therefore not guaranteed
(or expected) to be representative of galaxies at high redshift, and the
templates need to be extended into the UV and NIR for use with photometry from
modern multi-wavelength surveys.  Several of the recent photometric redshift
efforts address the first issue by iteratively adjusting the basis CK
templates based on fits to large broad-band photometric datasets \citep[ZEBRA;][]{ilbert:06, assef:07}.  While this technique is shown to
significantly improve the quality of the redshift estimates, it requires a
large spectroscopic redshift calibration sample and remains largely unproven at
$z\gtrsim1.5$.

We derive a new minimal template set based purely on stellar population synthesis models
that is designed for deep optical-NIR broad-band surveys and that requires no
optimization based on spectroscopic samples.  The template set is calculated
following the novel ``nonnegative matrix factorization'' (NMF) algorithm of
BR07, which essentially takes a large number, $N_\mathrm{in}$, of
synthetic models and computes a reduced set of $N_\mathrm{out}$ basis templates
that best reproduce a supplied broad-band photometric calibration catalog. 
The $N_\mathrm{out}$ basis templates are non-negative linear combinations of
the $N_\mathrm{in}$ models and can be considered to be the
``principal-component'' spectral templates of the calibration catalog.  For
example, BR07 compute $N_\mathrm{out}=5$ basis templates from a
list of $N_\mathrm{in}=485$ \cite{bc:03} models that efficiently reproduce a
large sample of photometric observations from the SDSS.  

While the SDSS offers precision photometric and spectroscopic information for a
spectacular number of galaxies, it is limited to the nearby universe -- which
means that
templates determined from (or optimized by) SDSS observations are
subject to similar uncertainties when extended to higher redshifts as those of
the CK templates described above.   Flux-limited spectroscopic samples
are now available up to $R\approx 24.5$
(e.g. DEEP2, \cite{deep2}; VVDS,
\cite{vvds}), but deep imaging surveys reach well beyond the limits
of these spectroscopic surveys.  In order to obtain a calibration sample that extends to faint magnitudes and
high redshifts, we turn to theoretical models of galaxy formation and
evolution that are complete to the extent that they reproduce observed galaxy
properties at high redshift.  Specifically, we obtain synthetic UBVRIzJHK
``photometry'' of galaxies in a 1 deg$^2$ lightcone \citep{momaf} created from
galaxies in the semi-analytic model (SAM) of \cite{lightcone_sam}, which is
based on the Millennium Simulation \citep{millennium}.  Synthetic spectral energy
distributions (SEDs) are generated from \cite{bc:03} models following the
non-trivial star formation histories of the galaxies in the semi-analytic
model.  While the models do not exactly reproduce the relative fractions of red
and blue galaxies at high redshift \citep{marchesini:07} and the simplified
treatment of dust obscuration is probably inadequate \citep{kitzbichler:07},
the models should contain a more representative sample of galaxy SEDs over the
broad redshift range $0<z\lesssim 4$ than purely local surveys.  For the calibration
set of the template optimization routine, we randomly select a subsample of
$10^4$  galaxies with $K_\mathrm{AB}<25$ from the lightcone catalog.  The
redshift and $K$ magnitude distributions of galaxies in the calibration sample
are shown in Figure \ref{fig:lc_dist}.

\begin{figure*}
\plotone{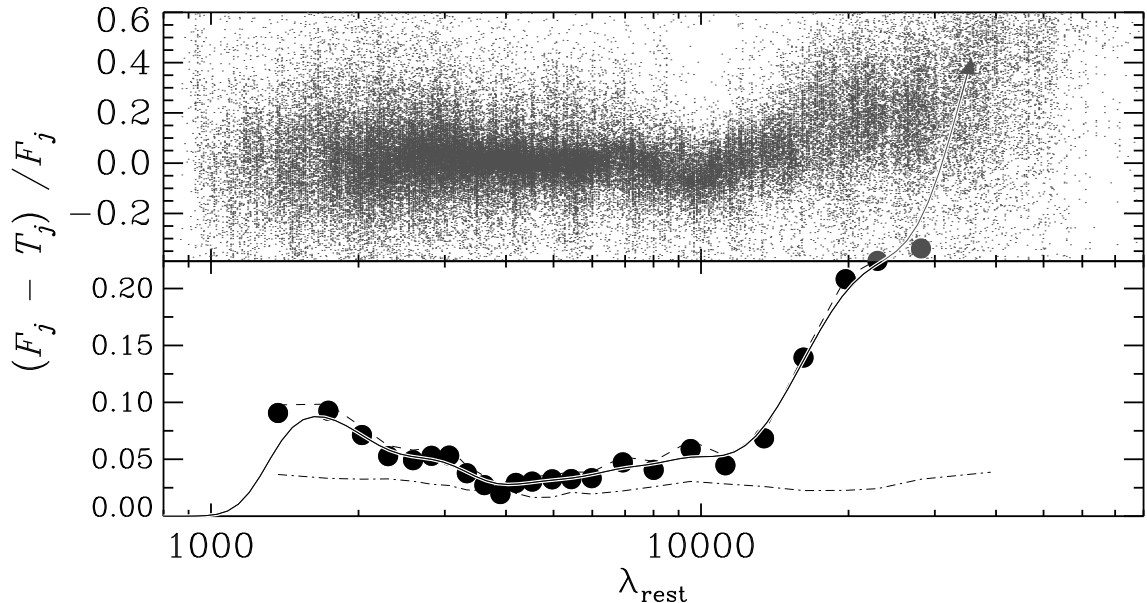}
\figcaption{\textit{top panel:} Normalized residuals $\left(\Delta F_j=
T_{z,j}-F_j\right)$ of redshift/template fits to the full GOODS-CDFS
photometric catalog \citep{wuyts:08}, shifted into the rest-frame using the
photometric redshifts of each source, estimated using the ``NMF'' templates
described in \S\ref{s:template_set}.  Only objects with a measured flux in all
bands are shown.  The dashed line indicates the median of the residual absolute values in wavelength
bins with widths such that each bin contains $\sim$2000 data points. 
\textit{bottom panel:} Binned residuals (absolute value) shown with the vertical scale expanded. 
The dashed line is the same as in the top panel.  The dot-dashed line is the
median \textit{photometric} error, $\sigma_j/F_j$, multiplied by a factor of
0.67 to scale the errors from 1-$\sigma$ (68\%) to 50\% confidence intervals. 
We subtract in quadrature the median photometric error from the median observed
residual in each bin to determine the contribution from ``template-error''
(filled circles).  The solid line is our adopted template error function, which
has been smoothed and extended into the NUV to avoid a discontinuity at the
Lyman break.  We extend the error function beyond the longest wavelength bin to have
$\sigma_{TE}=1.0$ at $\lambda=10\ \mu$m to account for the presence of dust
emission in the NIR that is not included in the templates.\label{fig:temp_err}}
\end{figure*}

For the input list of $N_\mathrm{in}$ models to the BR07 NMF
algorithm, we use the library of \pegase\ models \citep{pegase} described by
\cite{grazian:06}, which those authors use to obtain high-quality photometric
redshifts over $0<z<2$ in the GOODS-South field.  The library includes
$N_\mathrm{in}\sim3000$ models with ages between 1 Myr and 20 Gyr and having a
variety of star formation histories including exponentially decreasing star
formation rates characterized by the exponential decay rate, $\tau_*$, and
constant star formation models that are truncated.
Roughly half of the models in the library are constant star
formation models with additional reddening [$0.5\le E(B-V)\le1.1$] applied
using the extinction curve of \cite{calzetti:00}, which are designed to
represent young, dusty objects.  \pegase\ models provide a self-consistent
treatment of emission lines, which are not included in the \cite{bc:03}
models.  Though the synthetic photometry in the lightcone catalog does not
include emission lines, we use the \pegase\ template set with and without
emission lines and find slightly better results with the output templates when
emission lines are included.

The final $N_\mathrm{out}=5$ templates computed from the individual \pegase\
templates as fit to the lightcone catalog are shown in Figure
\ref{fig:show_templates}, along with the BR07 and CK
template sets for comparison.  The two NMF-derived template sets span
a larger range
of optical-NIR colors than the CK set, and the NMF templates
contain  more information in the NUV than the simple power-law
extrapolations of the CK set.  The higher spectral resolution of the
pure-model NMF templates is not likely to affect photometric redshift
estimates, but more realistic colors in the rest-UV should be important as
optical filters sample this portion of the SED at moderate redshift,
$z\gtrsim1$.  The primary difference between the BR07 template set
and the set derived here is the presence of the dusty (and old) template that
is the reddest template in the BR07 set.  Though there are models
with significant dust absorption in the \pegase\ library, the NMF algorithm
does not require a dusty template to fit the lightcone photometric catalog. 
This is likely due to the fact that the simple dust prescription in the
lightcone model is unable to produce any extremely dusty galaxies.  Therefore we add a dusty starburst model ($t=50\mathrm{\ Myr}, A_V=2.75$) to the set of 5 NMF-derived templates to compensate for the lack of dusty galaxies in the SAM calibration sample (Figure \ref{fig:show_templates}).  While the parameters chosen for this additional template are somewhat arbitrary, they are chosen such that the template fills in regions of the rest-frame color space not sampled by the 5 NMF-derived templates (Figure \ref{fig:show_templates}).  In \S\ref{s:application}
we compute photometric redshifts for a variety of publicly available optically-
and $K$-selected photometric survey fields and we compare the quality of
redshifts estimated using these three template sets.  

\subsection{The template error function}
\label{s:template_error}

Multi-wavelength surveys frequently sample rest-frame wavelengths from
the UV to the near-IR for galaxies at $z\lesssim 4$, and the quality of
the calibration of population synthesis models is not constant over that
full wavelength range. This can be caused by a number of factors, such
as (1) uncertainties in the stellar evolutionary tracks; (2)
transformation of the physical parameters of the models to observable
quantities; (3) variations in the dust extinction law; and (4)
stochastic spectral features that are simply not included in the models.
 For example, relevant to item (1) above, \cite{maraston:05} find that
short-lived thermally-pulsating asymptotic giant branch stars, which had
not been previously properly included in isochrone synthesis models, can
contribute significantly to the emergent NIR flux from stellar
populations younger than $\sim$1.5 Gyr.  Considering item (4), the
presence/strength of emission lines depends strongly on properties of
the ISM, which are only loosely coupled to the evolution of stars within
a galaxy.  For example, in the extreme case of Lyman-$\alpha$,
\cite{steidel:00} find equivalent widths ranging over 1--2 orders of
magnitude in emission and absorption for a relatively homogeneous sample
of Lyman-break galaxies at $z\sim3$.

No single generalized template set could hope to account for all of these
uncertainties, and it is therefore no surprise that corrections to a minimal
template set are required to optimize photometric redshifts for a given
photometric catalog \citep[e.g.][]{ilbert:06}.  As discussed in
\S\ref{s:template_set}, such optimization requires an extensive calibration
set of galaxies with spectroscopically-measured redshifts, which is often
unavailable or at best incomplete for deep imaging surveys.  Here we
derive a ``template error function'' that seeks to incorporate uncertainties
such as those mentioned above into the template fitting algorithm
(Eq. \ref{eq:chi2}).  Besides the calibration uncertainties described above, any set of individual templates will have difficulties reproducing the wide variety of star formation histories and dust extinction in galaxies.  To make matters worse, such properties not only vary among galaxies at a given cosmic epoch, but also vary systematically with time (or redshift).  Along with allowing for multiple linear combinations of individual templates, the template error function developed here helps to account for these variations.  The exact form of the template error function depends on the
chosen set of templates, but it is computed in such a way that it
generally applicable especially when no spectroscopic calibration sample is
available.

The template error function is derived in the following way. First,
photometric redshifts are determined with a uniform template error
function \citep[set at a constant 0.05 mag; see, e.g.][]{rudnick:01},
for the photometric catalog of the GOODS-CDFS field described by
\cite{wuyts:08}.  We use the CDFS because it provides the deepest survey
with extensive multi-wavelength coverage that includes the NIR IRAC
bands.\footnote{Since we wish to determine how well the \pegase\
template set matches observed data, we cannot use the synthetic
lightcone photometry as that would only illustrate differences between
SEDs produced by the \pegase\ and \cite{bc:03} population synthesis
codes.} Next, we calculate the residuals from the best-fitting model
spectra and de-redshift them into the rest-frame. These residuals are
shown in the top panel of Fig.\ \ref{fig:temp_err} (after several
iterations). The binned median absolute values of these residuals are
shown by the solid symbols in the bottom panel of Fig.\
\ref{fig:temp_err}, along with a smoothly varying function that is fit
to the solid symbols (dashed line). Finally, the template error function
(solid line) is created by subtracting in quadrature the (scaled)
photometric errors (indicated by the dot-dashed line) from this smoothly
varying function. The procedure is repeated until convergence is
reached.

The residuals in Figure \ref{fig:temp_err} are only shown for bands with
signal-to-noise (S/N) $>$ 10.  To test whether the derived error
function depends on the S/N of the flux measurements, we also compute
the template error for different limits 3 $>$ S/N $>$ 20.  The sizes of
the median residuals and photometric errors decrease as the S/N limit
increases, but we find that the quadratic difference of the two remains
mostly unchanged. The shape of the template error function roughly
follows what one might expect following the considerations enumerated
above:  the template error is lowest in the rest frame optical,
$\lambda=$3500\,\AA -- 9000\,\AA, where stellar isochrones are
well-calibrated; the template error is large in the UV where dust
extinction is strongest and most variable; and the template error
increases again in the NIR where the stellar isochrones are uncertain
\citep[e.g.][]{maraston:05} and where thermal dust emission and
stochastic PAH line features begin to appear at
$\lambda>3\mu\mathrm{m}$.

\subsection{Bayesian Prior}

\begin{figure}
\plotone{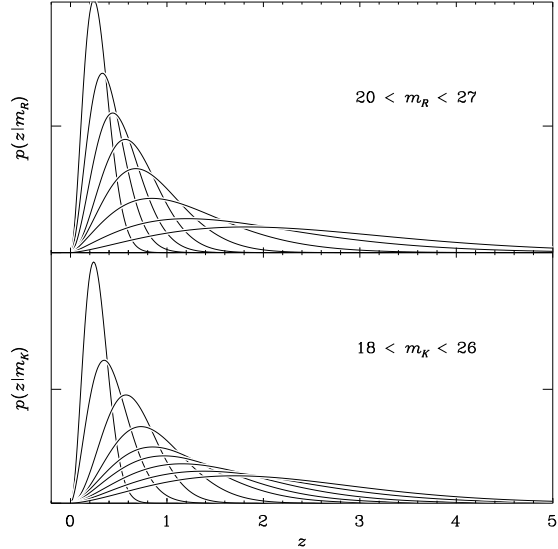}
\figcaption{Prior probabilities, $p(z|m_0)$, a function of observed $R$ and $K$
(AB) magnitudes.  The shape of the priors is given by the model redshift
distributions of galaxies in the lightcone catalog, normalized to
$\int p(z|m_0\ dz=1$.
  \label{fig:show_priors}}
\end{figure}

\begin{deluxetable*}{lclrc}
\tablehead{\colhead{Field} & \colhead{Area (arcmin$^2$)} & \colhead{Bands} & \colhead{K depth (3$\sigma$)} & \colhead{Ref.} }
\scriptsize
\tablewidth{0pt}\startdata
HDF-N\dotfill	 			&  $\sim$5 &  $UBVIJHK$ &  $\sim$24 & \markcite{hdfn}{Fern{\' a}ndez-Soto},  {Lanzetta}, \& {Yahil} (1999)  \\
HDF-S\dotfill 				&  4.5 &  $UBVIJHK+IRAC$&  26.2 & \markcite{labbe:03}{Labb{\' e}} {et~al.} (2003)  \\ 
MS-1054\dotfill			&   29 &  $UBVV_{606}I_{814}JHK+IRAC$ &  25.5 & \markcite{forster:06}{F{\"o}rster Schreiber} {et~al.} (2006) \\ 
MUSYC\dotfill 				&  400 &  $UBVRIzJHK+IRAC$ &  23.5 & \markcite{quadri:07}{Quadri} {et~al.} (2007), Marchesini et al. (in prep)  \\
GOODS-CDFS\dotfill 	&  138 &  $[U_{38}BVRI]_{\rm WFI}[BVIz]_{\rm ACS}JHK+IRAC$ &  24.9 & \markcite{wuyts:08}{Wuyts} {et~al.} (2008) \markcite{popesso:08}{Popesso} {et~al.} (2008) \\
\enddata \label{tab:fields}
\end{deluxetable*}

The template-fitting method of estimating photometric redshifts suffers from the
fact that template colors are frequently degenerate with redshift, such that the
redshift probability distributions can have multiple peaks over a broad range of
redshifts.  For example, relatively featureless blue SEDs can often be fit
equally well at $z=0$ and $z\sim3$ because the templates are unable to
distinguish blue colors redward of the Balmer and Lyman breaks, respectively. 
The degeneracies can sometimes
be broken by adding additional photometric bands (in
the previous example, adding IRAC photometry helps) or by
incorporating statistical methods to help choose between multiple probability
peaks at different redshifts.  

\cite{bpz} was the first to develop a Bayesian approach to estimating
photometric redshifts that includes the use of a Bayesian prior, which adds additional information besides the observed photometric
colors to help constrain the redshift estimates.  Following \cite{bpz} we adopt
an apparent magnitude prior, $p(z|m_0)$, which is the redshift distribution of
galaxies with apparent magnitude, $m_0$.  This is essentially a ``luminosity-function-times-volume'' prior that assigns a low probability to very low redshifts where the volume sampled is small and a similarly low probability of finding extremely bright galaxies at high redshift.  In contrast to \cite{bpz}, we do not include spectral (template) 
type in the prior because (1) our derived templates do not directly correspond
to individual galaxy spectral types; (2) we fit linear combinations of all 5
templates simultaneously; and (3) we do not want the prior to impose any color
restrictions as a function of redshift, the last point being most important. 
For example, the prior used by \cite{bpz} based on the HDF-N gives essentially
zero probability to red E/S0 spectral types at $z>2$, even though recent work
has shown that such galaxies are fairly common, at least in NIR-selected samples
\citep[e.g.][]{kriek:06}.

To determine the shape of the prior probability distribution,
$p(z|m_0)$, we again turn to the synthetic photometry of the SAM
lightcone catalog described in \S\ref{s:template_set} because this
problem is subject to many of the same completeness limitations of
observed samples as the template optimization routines.  In principal,
one could iteratively determine redshift distributions from observed
data and then recompute photometric redshifts using the distributions as
the prior, but observed samples are generally small at high redshift and
the iterative method is not guaranteed to converge to the truth.  Though
the synthetic models do not perfectly reproduce observed data (e.g.
Figure \ref{fig:lc_dist}), they should be able to reasonably estimate
$p(z|m_0)$ over $0<z \lesssim 4$, since in practice it is only the shape
of $p(z|m_0)$ that matters in a given $m_0$ bin and not the overall
normalization of the number of galaxies in that bin.  We adopt a
functional form of the prior,

\begin{equation}
p(z|m_{0,i})\propto z^{\gamma_i}\ \exp\left[-\left(\frac{z}{z_{0,i}}\right)^{\gamma_i}\right], \label{eq:prior_func}
\end{equation}

\noindent \citep{bpz} and fit the parameters, $\gamma_i$\ and $z_{0,i}$,
for the redshift distributions in each magnitude bin, $m_{0,i}$. 
Because the lightcones contain many galaxies at high redshift, the
functional fits do not extrapolate at high redshift, but rather they
ensure that the shape of the prior is smooth over the entire redshift
range.  Figure \ref{fig:show_priors} shows $p(z|m_0)$ for two selection
bands, $R$ and $K$, determined from the full 1 deg$^2$ lightcone catalog
of $\sim 10^6$ galaxies.

\begin{figure*}
\plotone{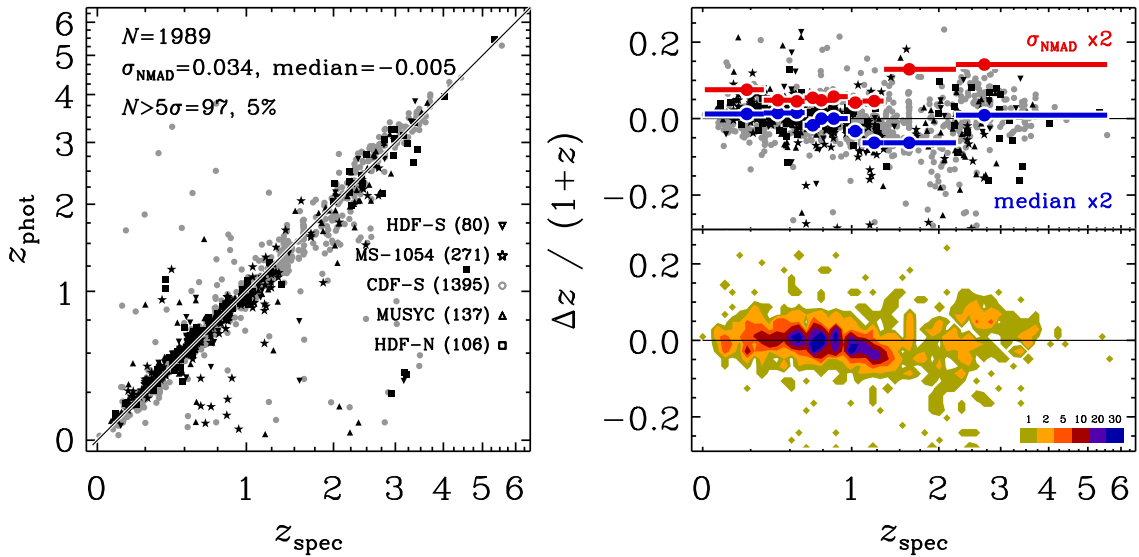}
\figcaption{\textit{left}: Spectroscopic vs. photometric redshifts
computed for 5 surveys with deep optical-NIR photometry, shown on a
``pseudo-log'' scale.  \textit{right, top}: Residuals, $\Delta z =
\zphot-\zspec$, as a function of $\zspec$.  The red datapoints indicate
$\sigma(z)$ scaled by a factor of two in bins that each contain $N=200$
sources and have widths indicated by the horizontal bars.  The blue
points indicate the median residual in the same bins, also scaled by a
factor of two to magnify the low-level systematic effects.
\textit{right, bottom}: Same as the top panel, but plotting the
``surface density'' of points rather than the points themselves to
highlight systematic effects.\label{fig:photz_specz}}
\end{figure*}

\subsection{Output redshifts and confidence intervals}

With the tabulated values of the prior, we can now compute the posterior
redshift probability distribution for each galaxy, given the galaxy's
observed colors, $C$, and apparent magnitude, $m_0$:

\begin{equation}
p(z|m_0,C) \propto p(z|C) p(z|m_0), \label{eq:prior1}
\end{equation}

\noindent \citep{bpz} where $p(z|C) =
\exp\left[-\chi^2\left(z\right)/2\right]$ is the likelihood computed
from the template fits (Eqs.\ref{eq:chi2}, \ref{eq:t_i}) over a
user-supplied redshift grid.  Given the posterior probabilities, the
code produces two redshift estimates, $z_p$ and $z_{mp}$, where $z_p$ is
simply the redshift where the probability is at its maximum and $z_{mp}$
is the value of the redshift marginalized over the posterior probability
distribution,

\begin{equation}
z_{mp} = \frac{\int z\ p(z|C,m_0) dz}{\int p(z|C,m_0)\ dz}. \label{eq:zm2}
\end{equation}

\noindent For a gaussian probability distribution, $z_p = z_{mp}$.  In
practice $z_{mp}$ smooths out some small-scale systematic errors
apparent in $\zphot$--$\zspec$ comparisons and $z_{mp}$ allows the use
of a coarse redshift grid to speed up the execution of the code without
significant loss of precision in the output redshift estimates.

We compute formal lower and upper confidence limits, $z_{lo}$ and
$z_{up}$, for a confidence level, $\alpha$, by integrating the posterior
probability distribution from the edges until the integrated probability
is equal to $\alpha/2$:
\begin{eqnarray}
\frac{\alpha}{2} & = & \int_0^{z_{lo}} p(z|C,m_0)\ dz, \\ \nonumber
\frac{\alpha}{2} & = & \int_{z_{up}}^\infty p(z|C,m_0)\ dz, \label{eq:confidence_limits}
\end{eqnarray}
where the limits $\left(0,\infty\right)$ are replaced in practice by
user-specified parameters, $\left(\textsc{zmin,zmax}\right)$, and 1-,
2-, and 3-$\sigma$ confidence limits are computed with $\alpha = 0.317,\
0.046,\ 0.003$, respectively.

\subsection{Software}
 
The algorithm is implemented in a public software program, dubbed
``\eazy'' (for ``Easy and Accurate Redshifts from Yale''). The
user-interface of \eazy\ was modeled on the popular HYPERZ code, but the
underlying code is written independently. \eazy\ is controlled through a
parameter file whose defaults should provide good redshifts for most
applications.  We provide the optimized template set, template error function, and priors described above as default inputs, but the code accepts any user-defined version of these files in simple ASCII formats.  The code is fast, taking about four minutes to run the
6300 galaxies in the \cite{wuyts:08} FIREWORKS GOODS-CDFS catalog with linear combinations of the six default templates (\S\ref{s:template_set}) on an Apple MacBook running a 2 Ghz Intel Core 2 Duo processor.  For a redshift test grid, $0 < z < 6;\ \Delta z=0.01\left(1+z\right)$, \eazy\ requires 12 s compared to 270 s for HYPERZ (without fitting reddening) to run the entire CDFS catalog when fitting only the single best-fit template of the six (but see Figure \ref{fig:params}).    
The \eazy\ package, along with installation instructions, example files
and a user's manual, can be obtained from \texttt{http://www.astro.yale.edu/eazy/}.

\section{Application}
\label{s:application}

We have combined a diverse set of public photometric catalogs to test
the quality of the photometric redshifts computed by \eazy.  In this
section, we compare photometric redshifts to a large sample of
spectroscopic redshifts.  We stress that
we do not use the spectroscopic sample to
calibrate the photometric redshifts explicitly, because, as discussed
earlier, our spectroscopic sample is not necessarily representative of
the full photometric samples.  Rather, we use the spectroscopic sample
to illustrate how certain systematic effects depend on features of the
code implemented as described in \S\ref{s:implementation}.

\subsection{Combined test sample}

To adequately test the code, we require photometric catalogs with broad
multi-wavelength coverage from the UV through the NIR to produce
unbiased photometric redshifts over a broad redshift range.  Deep NIR
photometry is essential for uncovering the complete population of
galaxies at $z>1.5$ \citep[e.g.][]{dokkum:06}, and is correspondingly
important for estimating photo-zs as the Balmer break shifts into the
NIR bands at these redshifts.  Table \ref{tab:fields} summarizes the
photometric data we use with \eazy\ to compute photometric redshifts\footnote{We provide $\zphot$ catalogs for the FIRES, MUSYC, and FIREWORKS surveys at \texttt{http://www.astro.yale.edu/eazy/} that supercede the $\zphot$ estimates provided by the catalog references listed in Table \ref{tab:fields}}.  All
of these catalogs provide $U$-band photometry necessary to break the
Lyman-break degeneracy between $z\sim0$ and $z\sim3$, and the catalogs
represent most of the deepest public NIR photometry available.  The
optical photometry for 4/5 of the photometric catalogs comes from the
\textit{Hubble Space Telescope} (HST).  We use the CDFS-GOODS catalog of
\cite{wuyts:08} which combines the deep HST GOODS photometry with
ground-based UBVRI photometry from the ESO Deep Public Survey
\citep{dps}.  The NIR $JHK$ photometry comes from a variety of
ground-based facilities. Additionally, the majority of the sample is
observed in the four \textit{IRAC} bands on the \textit{Spitzer Space
Telescope}.  Sources in the HDF-N are selected in the \textit{I}-band
(WFPC2-F814W), while sources in the other four fields are all selected
in $K$-band images.

Though the photometry in the fields listed in Table \ref{tab:fields}
reaches significantly deeper magnitudes than the practical spectroscopic
limit, these fields have been observed with extensive follow-up programs
that provide a large sample of spectroscopically-measured redshifts. We
have collected a sample of 1989 spectroscopic redshifts from references
listed in Table \ref{tab:fields} covering the full range $0 < \zspec
\lesssim 4$.  We use only the most reliable redshift quality flags when
they are available in the spectroscopic catalogs.  The spectroscopic
sample contains 334 galaxies at $\zspec > 1.5$ that have a variety of
spectral types, including Lyman break galaxies with blue rest-frame
colors \citep[LBGs;][]{steidel:03} and Distant Red Galaxies with quite
red rest-frame colors \citep[DRGs;][]{franx:03,kriek:06}.

\begin{figure*}
\plotone{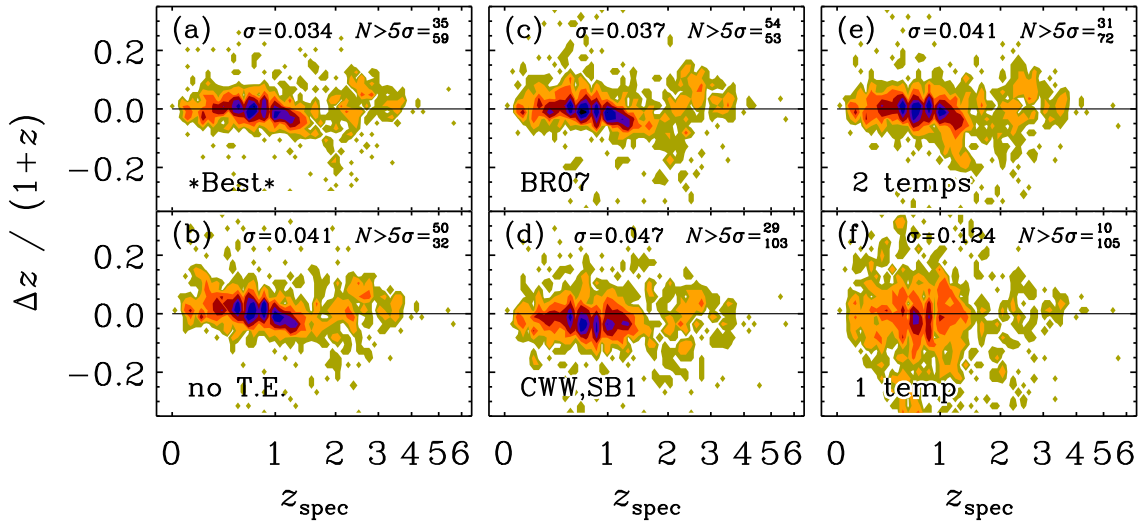}
\figcaption{Residuals, $\Delta z = \zphot-\zspec$, for different
combinations of input parameters.  The scatter, $\sigma$, and the number
of $5\sigma$ outliers are indicated in each panel with the outliers
divided into $N>5\sigma$ (superscript) and $N< -5\sigma$ (subscript). 
\textit{left, top}: identical to Figure \ref{fig:photz_specz}.  The
``best'' parameter set fits all of the \pegase\ templates simultaneously
(\S\ref{s:template_set}) and includes the template error function
(\S\ref{s:template_error});  \textit{left, bottom}: Template error not
used; \textit{center, top}: \texttt{kcorrect} templates from
\cite{blanton:07}; \textit{center, bottom}: Empirical templates from
\cite{cww} and \cite{kinney:96}; \textit{right, top}: Pairs of \pegase\
templates, rather than all simultaneously; \textit{right, bottom}:
Single \pegase\ templates. \label{fig:params}}
\end{figure*}

\subsection{Results with default parameters}

Figure \ref{fig:photz_specz} shows that $\zphot$ estimated by \eazy\
agrees remarkably well with $\zspec$ over the entire redshift range
covered by the spectroscopic sample. The same code parameters are used for
all fields, and \textit{no additional template or photometric optimizations are done} based on the $z_{\rm spec}$--$z_{\rm phot}$ comparison. We use the normalized median absolute
deviation ($\sigma_\textsc{nmad}$) of $\Delta z = \zphot-\zspec$ to
quantitatively assess the quality of the photometric redshifts, with
\begin{equation}
\sigma_\textsc{nmad} = 1.48 \times \mathrm{median}\left( \left| \frac{\Delta z-\mathrm{median}(\Delta z)}{1+\zspec} \right| \right) \label{eq:nmad}.
\end{equation}
With this definition, $\sigma_\textsc{nmad}$ is equal to the standard
deviation for a Gaussian distribution.  An advantage of this definition
is that it is less sensitive to outliers than the usual definition of
the standard deviation \citep[e.g.][]{ilbert:06}.  Hereafter we drop the
subscript for clarity ($\sigma=\sigma_\textsc{nmad}$).  The scatter,
$\sigma$, is nearly constant as a function of $\zspec$, with $\sigma =
0.034$ for the entire spectroscopic sample.  The scatter does increase
above $\zspec > 1.5$ where $\sigma = 0.075$.  Systematic deviations from
the $\zphot=\zspec$ line are very small at most redshifts, with the
exception that $\zphot$ underestimates $\zspec$ at $z=1.0-1.4$ by
$\sim$5\% (median).  Figure \ref{fig:photz_specz} shows a relatively
small number of sources that have estimated $\zphot$ very different from
$\zspec$.  These ``catastrophic outliers'' (defined here to have $\Delta
z/(1+z)>5\sigma$) make up 5\% of our spectroscopic sample.  We note that
we have not performed any cuts on the spectroscopic sample based on
signal-to-noise or coincidence with X-ray sources that could indicate
the presence of an AGN, both of which could contribute to a poor
estimate of $\zphot$.  In \S\ref{s:qz} we describe how the
catastrophic outlier fraction can be decreased using observables
computed from the $\zphot$ fit.  
 
\subsection{Effects of changing the default parameters}

We have computed $\zphot$ for the spectroscopic sample using different
combinations of input parameters to demonstrate the effects that the
features of the code implementation (\S\ref{s:implementation}) have on
the quality of the computed $\zphot$.  In a separate paper, E. N. Taylor et al. (in prep) present additional quantitative tests that show not only how the \eazy\ $\zphot$ estimates depend on changing input parameters such as the template set, but how the science results based on those redshifts---specifically the evolution of rest-frame colors and stellar masses of red galaxies over $0<z<2$---also vary systematically with different code inputs.
The residuals, $\Delta z$, for
the different $\zphot$ sets are shown in Figure \ref{fig:params}.  The
``best'' reference parameter set fits the 6 \pegase\ NMF templates
simultaneously and includes the template error function.  Figure
\ref{fig:params}b shows the residuals for fits that do not include the
template error function; the scatter is somewhat higher than when the
template error is used, and systematic effects appear as a function of
$\zspec$.  Galaxies at $z<1$ have $\zphot$ overestimated by 10\%.  The
number of catastrophic outliers is similar whether or not the template
error function is used, though when the template error function is not
used the majority of outliers have $\zphot >> \zspec$.  This is
potentially problematic for science applications that
would be adversely affected by bright, low-$z$ galaxies scattering
into high-$z$ samples (e.g. luminosity functions).  The number of
$5\sigma$ outliers is nearly constant in all of the panels of Figure
\ref{fig:params}, however the number of sources with $\Delta z /(1+z)$
greater than some fixed value, e.g 0.2, is significantly lower for the
best template/parameter set.

Figures \ref{fig:params}c,d show residuals for $\zphot$ computed using  the BR07 and empirical CK template sets, respectively.  The differences between fits using the \pegase\ and BR07 templates are small, though the systematic effects are somewhat worse when using the BR07 templates:  galaxies at $\zspec>1.5$ have $\zphot$ systematically low by $\Delta z\sim0.2$.  The scatter is significantly higher when using the empirical templates compared to either synthetic template set.   The $\zphot$ estimated with the empirical templates are also systematically underestimated at $0.5 < \zspec < 1.5$.  This effect has been observed previously in other photometric redshift studies and it has often been ``cured'' by correcting the templates based on the input photometry \citep[e.g.][]{zebra}.  However, we demonstrate here that our carefully-determined synthetic template set (Figure \ref{fig:photz_specz}) itself greatly reduces these systematic effects without requiring any additional corrections based on spectroscopic calibration samples.

Figures \ref{fig:params}e and \ref{fig:params}f show residuals for
$\zphot$ for $N_\mathrm{temp}=2\ \mathrm{and}\ 1$, respectively
(Eq. \ref{eq:t_i}), given the 6 templates of the \pegase\ NMF set.
The primary effect of fitting multiple templates simultaneously is a
striking reduction in $\sigma$.  This technique does not allow for the
simple spectral classification provided by single-template fits,
however the increased precision of the photometric redshifts should
allow more physical separations of photometric samples based on, for
example, rest frame colors.

\begin{figure}
\plotone{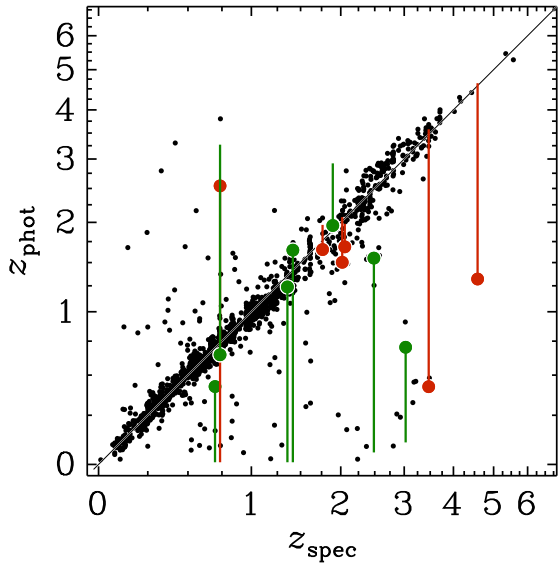}
\figcaption{Effect of the prior.  The sources where the prior moved the maximum likelihood redshift by $\left|\zprior-\zml\right| > 0.03 (1+\zspec)$ are indicated with a large point at $\zphot=\zprior$ and a tail extending to $\zphot=\zml$.  The cases where the prior improves $\zphot$ are shown in green, and the cases where the prior incorrectly pushes $\zphot$ away from $\zspec$ are shown in red.\label{fig:show_prior}}
\end{figure}

\begin{figure}
\plotone{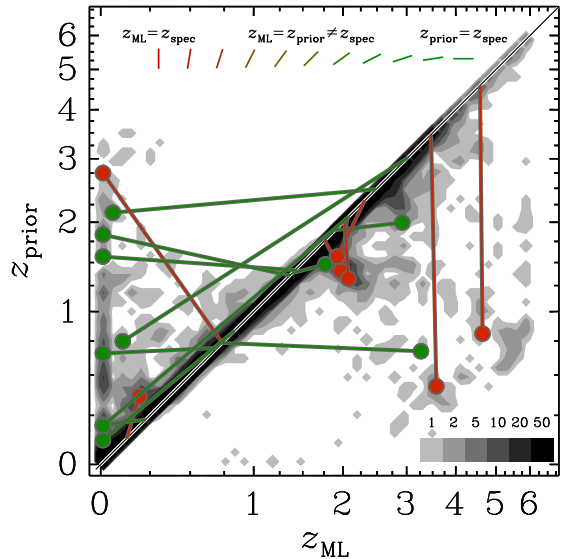}
\figcaption{Maximum likelihood $\zphot$ vs. $\zphot$ after including the prior (Eq. \ref{eq:prior1} for the full photometric sample of galaxies from the surveys listed in Table \ref{tab:fields}.  Sources with measured $\zspec$ and where $\left|\zprior-\zml\right| > 0.03 (1+\zspec)$ are indicated, as in Figure \ref{fig:show_prior}.  Here, the tails point from $\left[x,\ y\right]=$ $\left[\zml,\ \zprior\right]$ to $\left[\zspec,\ \zspec\right]$.  Therefore, horizontally oriented tails indicate cases where the prior improves $\zphot$.  Cases where $\zml\sim \zprior$ but both are different than $\zspec$, which would lie along the one-to-one line on the figure, are not shown.\label{fig:zphot_zphot}}
\end{figure}

Figure \ref{fig:show_prior} shows how the incorporation of the redshift
prior affects the $\zphot$ estimates.  While the prior generally
improves $\zphot$, it does not efficiently discriminate between multiple
probability peaks in all cases.  For example, there is a handful of
galaxies in the spectroscopic sample at $\zspec\sim3$ but with
$\zphot\sim0.2$ resulting from the degeneracy between fitting the Balmer
break at low redshift and the Lyman break at high redshift.  The prior
breaks this degeneracy for the single case indicated, but does not work
for the remaining galaxies with discrepant $\zphot$.  The prior does,
however, affect more sources with similar degeneracies from the full
photometric sample (without $\zspec$).  Figure \ref{fig:zphot_zphot}
indicates that there are ``clouds'' of sources in $\zml-\zprior$ space
that follow the behavior indicated by only one or two sources with
measured $\zspec$.  For example, there is a cloud of sources with
$\zml\sim0.2$ and $\zprior\sim3$.  The galaxies in this cloud likely
have similar colors that all result in the same redshift degeneracy of
the template fits \citep[see, e.g.][]{oyaizu:07}.  If the single source
with a measured $\zspec$ is representative of this group, than $\zprior$
is likely closer to the true redshift for all of the sources in this
group.  The opposite could be true for the group of sources at
$\zml>3.5$ and $\zprior<1$, where the two sources with $\zspec$
indicates that $\zml$ is likely a better estimate of the true redshift
than $\zprior$.  Nearly all of the ``clouds'' in Figure
\ref{fig:zphot_zphot} have one or two counterparts with measured
$\zspec$, so such a figure could be used to choose the optimal $\zphot$
estimate for all of the sources in the full photometric sample.

\subsection{Comparison to neural network redshifts}

If our assumption that the synthetic photometry of the lightcone catalogs is representative of true galaxy photometry over $0<z<4$ (\S\ref{s:template_set}), then the synthetic photometry could perhaps be used to train a neural network that could estimate photometric redshifts independently of our template-fitting approach.  We use the ANN$z$ code \citep{annz} to train a committee of five 11:10:10:1 neural networks directly on 11-band ($UBVRIzJHK+IRAC1,2$) lightcone catalog.  The photometric redshifts estimated for the validation sample (a random subset of the lightcone catalog) have $\sigma\sim0.03$, indicating that the network training works reasonably well.  We test the neural network on the observed photometry of the MUSYC HDFS-1 field, which contains 114 spectroscopic redshifts over $0<z<3$ and whose filter transmission curves are identical to those used to compute the synthetic lightcone photometry.  

The photometric redshifts estimated by (\eazy, ANN$z$) have $\sigma=(0.046, 0.100)$ for the full MUSYC-HDFS spectroscopic sample and $\sigma=(0.075, 0.105)$ for 18 galaxies at $\zspec>2$.  Along with the increased scatter, the ANN$z$ photometric redshifts show systematic errors that are not seen in the \eazy\ redshifts.  These discrepancies are likely caused by the fact that the neural network technique depends more critically on the assumption that the training catalog has identical properties to the full data catalog.  For example, \cite{annz} point out that ANN$z$ is able to account for the internal redenning of SDSS galaxies in their training and validation samples, but if the dust model of the lightcones is incorrect, the neural networks trained on them will have systematic problems matching real observations.  These sorts of problems also affect the template-fitting approach, though our additional dusty template and the template error function are designed to address such systematic uncertainties.

\section{Reliability of photometric redshifts}
\label{s:reliability}

\subsection{Confidence intervals}

We can assess how well the confidence intervals computed by Eq.
\ref{eq:confidence_limits} reflect the $\zphot$ uncertainties by
observing how often the measured $\zspec$ falls within the interval. 
For example, 41\% of sources have $\zspec$ outside of the 68\%
confidence interval, while 32\% are expected .  In general, however,
$\zspec$ lies close to the edge of the confidence intervals, and
expanding the 68\% confidence interval by a factor of only
$0.01\times(1+\zspec)$ decreases the number of discrepant sources to
29\%.  We find that the probability distributions, $p(z|C,m_0)$, and therefore the confidence intervals computed by Eq.
\ref{eq:confidence_limits} for individual sources, are consistent with
the $\zphot$ distributions of Monte-Carlo simulations in which we measure $\zphot$ after perturbing the
photometric fluxes within their associated uncertainties. 
 In general, we conclude that the confidence intervals provide a
reasonable representation of the uncertainties of the $\zphot$
estimates.
 
\subsection{Reliability parameter}
\label{s:qz}

\begin{figure}
\plotone{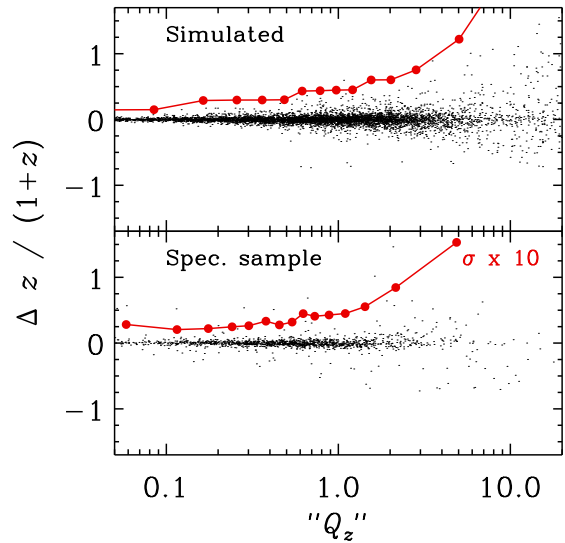}
\figcaption{Redshift residuals as a function of the redshift quality parameter, $Q_z$ (Eq. \ref{eq:qz}).  The red line indicates $\sigma(Q_z)$, scaled by a factor of 10 so that variations are visible on the plot scale. \label{fig:qz}}
\end{figure}

There is a small number of sources where $\zspec$ lies well outside even the 99\% confidence intervals.  These sources usually have sharply-peaked probability distributions, so no alternate definition of Eq. \ref{eq:confidence_limits} \citep[e.g.][]{fernandez:02} or Monte-Carlo simulations could significantly improve the confidence intervals.  Catastrophic outliers can be caused by (a combination of) a number of factors: (1) intrinsic SEDs that are not well-reproduced by the template set, (2) degeneracies in color-$z$ space that result in multiple peaks in $p(z)$--especially for very blue galaxies with featureless SEDS--(3) one or more anomalous photometric measurements, or (4) simply that the spectroscopic source was incorrectly matched to a photometric source or that the redshift was misidentified in the spectrum.  Quantitative features of the $\zphot$ fit can be used to identify catastrophic outliers caused by one or more of the problems described above.  For example, \cite{bpz} defines a parameter, $p_{\Delta z}$, that represents the fraction of the total integrated probability that lies within $\pm\Delta z$ of the $\zphot$ estimate, and is designed to identify sources that have broad and/or multi-modal probability distributions.   \cite{mobasher:07} find that the $\zphot$ scatter is an increasing function of a parameter, $D95$ that is defined as the ratio of the 95\% confidence interval to (1+$\zphot$).  Here we define a parameter, $Q_z$, that is a hybrid of the parameters proposed by \cite{bpz} and \cite{mobasher:07}, and also includes the $\chi^2$ of the template fit:  
\begin{equation}
Q_z = \frac{\chi^2}{N_{\rm filt}-3} \frac{z_{up}^{99}-z_{lo}^{99}}{p_{\Delta z=0.2}}. \label{eq:qz}
\end{equation}
The inclusion of $\chi^2$ should allow us to address the ``catastrophic'' cases (1-3) above.  Figure \ref{fig:qz} shows the $\zphot$ residuals and $\sigma$ as a function of $Q_z$.  We show $Q_z$--$\sigma$ for the spectroscopic sample and also for a simulated sample following \cite{bpz}, where we fit $\zphot$ for each source in the CDFS photometric catalog; set the template colors of the fit at $\zphot$ to be the new photometric colors; add photometric scatter following the photometric errors; and finally refit $\zphot$ for the new synthetic photometry.  The $\zphot$ scatter increases sharply above $Q_z=2-3$ in both the synthetic and observed spectroscopic samples.  The $5-\sigma$ outlier fraction at $Q_z>2 (3)$ is 0.3 (0.4), so quality cuts based on $Q_z$ can eliminate a large fraction of the outliers at the expense of a small number of satisfactory $\zphot$ estimates.  We have verified that quality cuts based on $Q_z$ are independent of redshift, that is, cutting on $Q_z>3$ does not preferentially remove only high-$z$ sources.

\subsection{The false security of $z_{\rm phot}$ -- $z_{\rm spec}$ plots}

We have emphasized that \eazy\ has been designed to estimate photometric
redshifts of galaxies in deep photometric surveys that lack
representative calibration samples with measured spectroscopic
redshifts.  Such a situation will be the case for the latest generation
of large NIR surveys that will reach $K\sim25$ \citep[e.g. UKIDSS;][]{ukidss} and probe galaxies with $L^\star$ luminosities at
$z\sim3.5$ \citep[e.g.][]{marchesini:07a}.  Although \eazy\ (and other
codes) performs very well in $z_{\rm phot}$ -- $z_{\rm spec}$ plots,
many systematic effects are ``hidden'' in such diagrams. This is
implicit in Fig.\ \ref{fig:zphot_zphot}: this Figure demonstrates that
there are large groups of sources whose photometric redshifts are very
sensitive to the details of the optimization routine, but that this
behavior is only ``sampled'' by, at-best, one or two sources with
measured $\zspec$.  A single outlier, such as the object at
($\zspec=4.5$, $\zphot=1.25$), might not appear noteworthy in a plot
like Figure \ref{fig:show_prior}, but it could represent a large number
of objects.

\begin{figure}
\epsscale{0.98}
\plotone{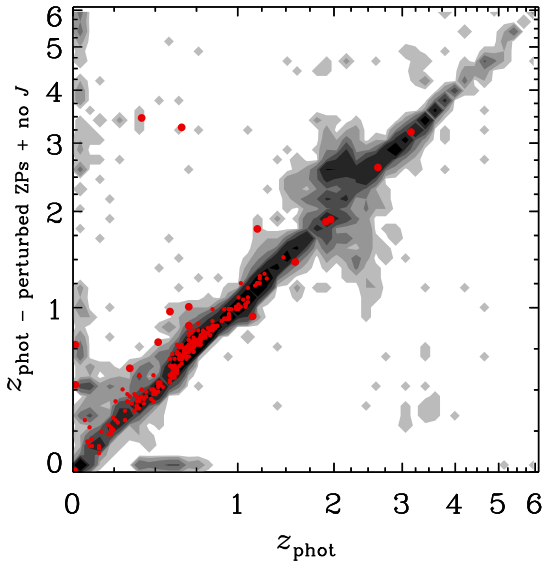}
\figcaption{Comparison of $\zphot$ computed before and after adding
random zeropoint errors of up to $\sim$5\% to the CDFS photometric
catalog of \cite{wuyts:08}.  The full photometric sample is shown in the
greyscale contours, while the red points indicate galaxies with
spectroscopic redshifts measured by the VVDS \citep{vvds}. The contour
levels are the same as those in Figure \ref{fig:zphot_zphot}.
\label{fig:perturb}}
\end{figure}

We explore this effect further by comparing $\zphot$ computed for the
full CDFS catalog of \cite{wuyts:08} and for a perturbed
version of the same catalog.  For the perturbed catalog, we add random
zeropoint offsets to each of the photometric bands with a maximum
offset of 5\%, and we remove the $J$-band from the $\zphot$ fit.  The
$\zphot$ for the normal and perturbed catalogs are shown in Figure
\ref{fig:perturb}.  The small zeropoint offsets cause systematic
effects visible as kinks in Figure \ref{fig:perturb} between
$0<\zphot<1$.  Without the $J$-band, the template fits cannot
efficiently isolate the Balmer break at $1.5<z<2.5$, and the
degeneracy of the $\zphot$ fits at these redshifts are visible in
Figure \ref{fig:perturb}.  Now we consider how these systematic
effects are traced by the subsample of galaxies from a flux-limited
spectroscopic survey, in this case the $I<24$ spectroscopic survey of
the VVDS \citep{vvds}.  At redshifts where the spectroscopic survey is
complete, here $z\lesssim1.5$, the spectroscopic sample effectively
traces the systematic effects of the full photometric sample and the
sources with spectroscopic redshifts can be used to calibrate the
photometric redshift algorithms.  There are a few galaxies at $z>2$ in
the spectroscopic sample, but their $\zphot$ are nearly identical from
both the normal and perturbed photometry.  One would not be able to
distinguish between the two catalogs---or similarly between two
separate photometric redshift algorithms \citep{hildebrandt:08}---if
each produces the same $\zphot$ estimates, even if the $\zphot$
estimates suffer from serious systematic uncertainties for sources
beyond the flux limits of the spectroscopic comparison sample.

\begin{figure}
\plotone{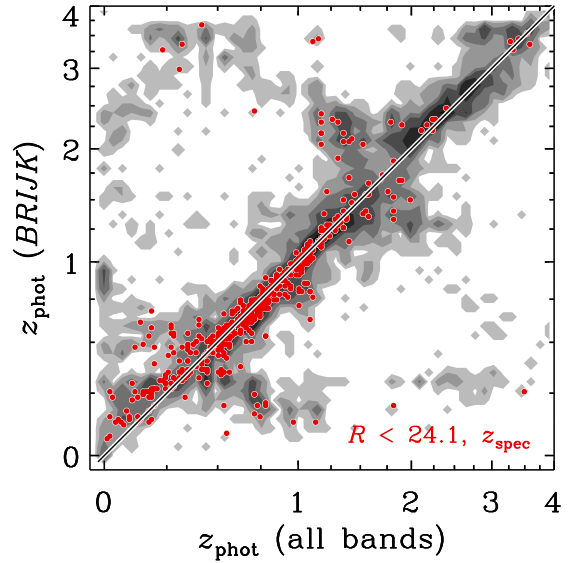}
\figcaption{Comparison of $z_{\rm phot}$ computed for the full CDFS filter set
and for a subset of the filters, $BRIJK$ \citep[the filter set used in the POWIR survey;][]{conselice:07}. CDFS sources with measured
$z_{\rm spec}$ and $R<24.1$ corresponding to the limit of the DEEP2
spectroscopic limit are shown in the large red points.
\label{fig:powir}}
\end{figure}

Similar effects are demonstrated in Fig.\ \ref{fig:powir}. Here, we did
not perturb the zeropoints, but mimicked the filter set of the POWIR
survey \citep{conselice:07}, which is one of the largest
$K$-selected surveys to date.  Again, the spectroscopic sample (in this
case, redshifts down to the DEEP2 limit $R<24.1$) would suggest
reasonably robust redshifts for either the full CDFS ($\sigma=0.037$) or
partial POWIR ($\sigma=0.048$) filter sets at $1 < z < 2$.  Considering
the entire CDFS photometric sample however, it is apparent that there
are significant discrepancies between the two filter combinations at $z
> 1.25$.  These degeneracies have important implications for the
interpretation of high-$z$ galaxy samples, for example surface densities
in two bins, $1.5 < z < 2$ and $2 < z < 2.5$, differ by more than a
factor of 2 depending on the filters used to estimate $z_{\rm phot}$. 
These discrepancies would likely be more pronounced in surveys with limited filter coverage that are also significantly shallower than the CDFS photometry.

\section{Summary}
\label{s:summary}
This paper presents a new photometric redshift code, dubbed ``\eazy''.
The philosophy of the code is different from other recent photometric redshift codes, in that it
does not aim to minimize the scatter in $z_{\rm phot}$ -- $z_{\rm spec}$
comparisons. This type of minimization works well if the spectroscopic
sample is a random subset of the photometric sample, but may lead to
erroneous results if the spectroscopic sample is biased. In $K$- or
IRAC-selected samples the vast majority of objects is very faint in the
observer's optical \citep[see, e.g.][]{dokkum:06} and the subset of
galaxies with a spectroscopic redshift is highly a-typical.  We therefore develop a new template set (\S\ref{s:template_set}) based on synthetic photometry of galaxies in a semi-analytic model that is essentially complete at redshifts significantly beyond the reach of current spectroscopic surveys.  Furthermore, we introduce a ``template error function'' (\S\ref{s:template_error}) that accounts for both random and systematic differences between observed photometry and the templates and minimizes systematic errors in $\zphot$ without the need to optimize either the templates or the photometry based on a spectroscopic calibration sample.  The template set and template error function provided here are intended to be generally applicable to NIR-selected samples.  With these default parameters and without further optimization, we find that the scatter in $z_{\rm phot}$--$z_{\rm
spec}$ diagrams ($\sigma=0.034$) is at least as low as achieved by other methods and systematic errors are minimal over the full range $0 < z < 4$ (\S\ref{s:application}).

The reliability of the uncertainties in the redshifts is almost as
important as the reliability of the redshifts themselves. The
uncertainties that our code provides behave well for galaxies with a
spectroscopic redshift (in the sense that the confidence intervals
correctly describe the deviations from the true redshifts), but we
cannot test the behavior in the same way for the majority of objects
without a spectroscopic redshift.  Assessing the reliability of photometric redshifts for a sample in which the spectroscopic redshifts are not a representative sample of the full photometric sample can be misleading (\S\ref{s:reliability}).  This is especially true when only limited filter coverage is available and large systematic effects can be present that are not clearly traced by the spectroscopic sample.  In practice, we expect that the
``$Q_z$'' parameter (\S\ref{s:qz}) gives a reasonable indication of the robustness of a
redshift.

Progress in the study and interpretation of faint galaxy samples is
currently limited by the uncertainties in photometric redshift
estimates. What is sorely needed is a {\em complete} set of faint,
$K$-selected galaxies with reliable redshifts, so that photometric
redshifts can be better calibrated. At relatively bright $K$-magnitudes
progress can be expected from near-IR spectroscopy, particularly with
multi-object near-IR spectrographs. Initial results of such programs
already point to problems with some photometric redshift estimates
\citep{kriek:08}, although we note that the Kriek et al.\ sample is
well-fit by \eazy. Deeper samples may be obtained through different
methods. A program is underway with NEWFIRM on the Kitt Peak 4m
telescope to obtain medium band photometry for a complete sample of
$\sim 10^5$ galaxies with $K\leq 21.5$, and this should provide
much-needed tests of the broad-band results reported in the literature.

\acknowledgements
The authors are grateful to Gabriella de Lucia and Jeremy Blaizot for
computing a large set of model galaxy magnitudes for
the filters used in this paper and constructing a set of light-cones
specifically designed for our study, which are a critical
component of this paper. We also thank Stijn Wuyts, Ned Taylor, Rik Williams, and Ryan Quadri
for extensive tests of early versions of the code, and Gregory Rudnick and Marijn Franx
for valuable suggestions and comments.

\bibliographystyle{apj}
\bibliography{eazy}

\end{document}